\documentclass[useAMS,usenatbib]{mnras}
\usepackage{graphicx}
\usepackage{txfonts}
\usepackage{amssymb}
\usepackage{amsfonts}
\usepackage{times}
\usepackage{hyperref}

\title[Effect of PNG on Galaxy Clusters Scaling Relations]{Effect of Priomordial non-Gaussianities on Galaxy Clusters Scaling Relations}
 
\author[A. M. M. Trindade and Antonio da Silva]{A. M. M. Trindade$^{1,2}$\thanks{E-mail:
arlindo.trindade@gmail.com ; Arlindo.Trindade@astro.up.pt (ANO)} and Antonio 
da Silva$^{3,4}$\thanks{E-mail:asilva@astro.up.pt}\\
$^{1}$Instituto de Astrof\'{\i}sica e Ci\^encias do Espa\c{c}o, Universidade do 
Porto, CAUP, Rua das Estrelas, PT4150-762 Porto, Portugal \\
$^{2}$Departamento de F\'{\i}sica e Astronom\'{\i}a, Faculdade de Ci\^encias, 
Universidade do Porto, Rua do  Campo Alegre 687, PT4169-007 Porto, Portugal\\
$^{3}$Instituto de Astrof\'{\i}sica e Ci\^encias do Espa\c{c}o, Universidade de 
Lisboa, Faculdade de Ci\^encias, Campo Grande, PT1749-016 Lisboa, Portugal\\
$^{4}$Departamento de F\'{\i}sica, Faculdade de Ci\^encias, Universidade de 
Lisboa, Faculdade de Ci\^encias, Campo Grande, PT1749-016 Lisboa, Portugal
 }

\date{Accepted XXX. Received YYY; in original form ZZZ}

\begin{document}
\maketitle

\begin{abstract}
Galaxy clusters are a valuable source of cosmological information. Their formation and evolution depends on the underlying cosmology and on the statistical nature of the primordial density fluctuations. In this work we investigate the impact of primordial non-gaussianities (PNG) on the scaling properties of galaxy clusters. We performed a series of cosmological hydrodynamic $N$-body simulations featuring adiabatic gas physics and different levels of non-Gaussian initial conditions within the $\Lambda {\rm CDM}$ framework. We focus on the $T-M$, $S-M$, $Y-M$ and $Y_{X}-M$ scalings relating the total cluster mass with temperature, entropy and SZ cluster integrated pressure that reflect the thermodynamical state of the intra-cluster medium. Our results show that PNG have an impact on cluster scalings laws. The mass power-law indexes of the scalings are almost unaffected by the existence of PNG but the amplitude and redshift evolution of their normalizations are clearly affected. The effect is stronger for the evolution of the $Y-M$ and $Y_{X}-M$ normalizations, which change by as much as 22\% and 16\% when $f_{\rm NL}$ varies from $-500$ to $500$, respectively. These results are consistent with the view that positive/negative $f_{\rm NL}$ affect cluster profiles due to an increase/decrease of cluster concentrations. At low values of $f_{\rm NL}$, as suggested by present Planck constraints on a scale invariant $f_{\rm NL}$, the impact on the scalings normalizations is only a few percent, which is small when compared with the effect of additional gas physics and other cosmological effects such as dark energy. However if $f_{\rm NL}$ is in fact a scale dependent parameter, PNG may have larger positive/negative amplitudes at clusters scales and therefore our results suggest that PNG should be taken into account when galaxy cluster data is used to infer cosmological parameters or to asses the constraining power of future cluster surveys.

\end{abstract}

\begin{keywords}
Cosmology: large-scale structure of Universe; Cosmology: cosmological parameter
s; Methods: numerical; Galaxies: clusters
\end{keywords}

\section{Introduction \label{intro}}

The Inflationary paradigm has become the widely accepted mechanism responsible for
the generation of the primordial density perturbations that seeded the observed 
Large-Scale Structure (LSS) of the Universe. 
An important prediction of the simplest, single field, slow-roll inflationary standard theory 
is the generation of nearly gaussian distributed primordial density perturbations 
(see e.g. \citealt{2008JCAP...08..031S,2003JHEP...05..013M,2005PhRvL..95l1302L,2005JCAP...06..003S,2007PhRvD..76h3004S}). 
Present Cosmic Microwave Background (CMB), e.g. \citealt{2011ApJS..192...18K},  and 
large-scale structure, e.g. \citealt{2008JCAP...08..031S,2014A&A...571A..16P}, observations 
have not ruled out this prediction (in fact they support it). 
However, in more sophisticated inflationary models, where the conditions of the standard 
single-field slow-roll inflation fail, a significant 
and potentially observable deviation from Gaussianity may be produced. 

Extensive work has been developed to try
to detect and constraint primordial non-Gaussianities (PNG) 
using a wide range of cosmological probes. 
Such detection would considerably decrease the number
of viable inflationary models and would also  provide valuable 
insights on key physical processes that took place in the early Universe.

The three-point statistics of the CMB temperature anisotropies have been 
the prefered tool to try to constraint the level primordial non-Gaussianities.
In recent years, many attempts  have been made to motivate and use 
other cosmological probes for the same purpose. This includes using 
statistical properties of large-scale structure, namely the bispectrum and/or 
trispectrum of galaxy distributions 
(e.g. \citealt{2007PhRvD..76h3004S,2008ApJ...677L..77M}, 
\citealt{2012MNRAS.422.2854G}) and weak-lensing observations 
(e.g. \citealt{2012MNRAS.421..797S,2012MNRAS.426.2870H}), 
as well as CMB-LSS \citep{2012arXiv1205.0563T} and CMB-21cm line \citep{2012PhRvD..85d3518T} cross-correlations. 
The evolution with time of the abundance of both massive collapsed objects, 
such as galaxy clusters, and large voids has also been presented in the literature as 
an independent powerful method, to constrain cosmology and 
in particular non-Gaussian models (see e.g \citealt{2000ApJ...541...10M,2000MNRAS.311..781R} and \citealt{2009JCAP...01..010K,2009MNRAS.399.1482L,2011PhRvD..83b3521D,2012arXiv1204.2726S}).

The wide diversity of inflationary models available in the literature 
predict different levels of deviations from Gaussian initial conditions 
in the primordial density spectrum. 
Depending on the underlying physical mechanism responsible for 
the generation of non-Gaussianities, different triangular configurations 
(shapes) arise. 
There are broadly four classes of triangular shapes or, equivalently, 
four different bispectrum parametrizations: 
Local, Equilateral, Folded and Orthogonal.  
From these, the most studied are primordial non-Gaussianities of the Local type, 
 which are usually expressed in terms of the gauge-invariant Bardeen's potential, 
 written as a Taylor expansion around a auxiliary isotropic Gaussian random 
 field $\phi$ as (see e.g. \citealt{2007JCAP...03..005C}),

\begin{equation}
\label{PNG-def}
\Phi\left(\mathbf{x}\right) = \phi\left(\mathbf{x}\right) + f_{NL} \left( \phi^{2}\left(\mathbf{x}\right) - 
\langle\phi^{2}\left(\mathbf{x}\right)\rangle \right)\,,
\end{equation}
where $f_{NL}$ is a, scale-independent, non-linear parameter that 
controls the level of deviation from Gaussianity. 
The tightest constrain to date on the non-linear parameter $f_{NL}$ for this parameterization was achieved by the Planck collaboration, by measuring the three-point statistics of the CMB temperature anisotropies,  
$f_{NL} = 2.7 \pm 5.8 \; \left(68\%\; \mathnormal{C.L.}\right)$ \citealt{2014A&A...571A..24P} (  and more recently $f_{NL} = 2.5 \pm 5.7 \; \left(68\%\; \mathnormal{C.L.}\right)$ \citealt{2015arXiv150201592P}). 
This result is consistent with primordial Gaussian intial conditions assuming 
that $f_{NL}$ is a scale independent parameter. 
However, if  $f_{NL}$ varies with scale one may expect a different level of 
non-gaussianities on scales smaller than the CMB scales probed by Planck.

On galaxy cluster scales, primordial non-gaussianities are expected to influence the mass and redshift distribution of cluster abundances. According to \citealt{2012MNRAS.424.1442T,2013MNRAS.435..782T}, falling to take in to account the effect of PNG on clusters abundaces, can led to biases in the estimation of cosmological parameters when clusters counts are used as cosmological probes. 
Accurate predictions on how PNG afects clusters abundances, require detailed knowledge about the underlying mass function of cluster haloes as well as undertanding the way the total cluster mass relates to baryon observables. 

The impact of primordial non-gaussianities on the cluster halo mass function has been extensively investigated using a combination of analytical (see e.g. \citealt{2000ApJ...541...10M,2000MNRAS.311..781R,2008JCAP...04..014L,
maggiore:2010,damico:2011,2012JCAP...02..002A,daloisio:2013,lim:2014}) and numerical N-body (dark matter only) simulation (see e.g. \citealt{kang:2007,grossi:2007,grossi:2009,desjacques:2009,Pillepich:2010,2010JCAP...10..022W,smith:2011,2012JCAP...03..002W}) methods.   

The study of the effect of PNG on cluster baryon observables is hard to model analytically.
Hydrodynamic N-body simulations (that model both dark matter and baryons) are the most appropriate 
tool to follow the evolution of the complex baryon physics acting on inter-galactic (IGM) and 
intra-cluster medium (ICM) scales during the non-linear evolution of cosmological structure. 
A first study of the impact of primordial non-gaussianities on structure formation using hydrodynamic N-body techniques was carried out by \citealt{maio:2011} that modeled gas chemistry and a number of other gas physical processes in their simulations to study early gas properties, star formation, metal enrichment and the evolution of stellar populations. 
Latter, \citealt{zhao:2013}, also applied hydrodynamic simulations with chemistry and radiative gas physics to study the formation and evolution of galaxies within the PNG framework.   
More recently \citealt{pace:2014} carried out PNG hydrodynamic simulations, including cooling, star formation, stellar evolution and metal pollution from stellar populations, to study the Sunyaev-Zeld'ovich (SZ) signal, due to the inverse Compton scattering of CMB photons by ionized gas, in galaxy clusters and filamentary structures

Galaxy cluster number counts, e.g. from X-rays or SZ cluster surveys, are known to be a 
most promising method to constrain deviations from primordial gaussianity at cluster scales 
(see eg \citealt{sartoris:2010, roncarelli:2010, Pillepich:2012, mak:2012, khedekar:2013} for 
several cluster survey forecasts). 
These estimates critically rely on assumptions about the state of the ICM gas atmospheres and on 
the way their observed properties link with the total cluster mass.  
The link is usually expressed via galaxy cluster scaling relations that allow to convert mass function estimates into observed number counts.
These studies often assume hydrostatic equilibrium, spherical symmetry and the 
self-similar model for clusters \citep{kaiser:1986, kravtsov:2012}. 
More sophisticated approaches rely on galaxy cluster scaling relations derived from hydrodynamic or N-body simulations, calibrated by observations, 
that do not include primordial non-gaussianities (see eg \citealt{mak:2012, roncarelli:2010}).
This procedure is clearly not ideal given that non-gaussianities are known to influence the internal 
structure of clusters \citep{smith:2011, dizgah:2013} and therefore they may  
cause significant changes in the slope and normalization of galaxy cluster scalings. 
The study of the impact of PNG on cluster scaling relations is also essential for an accurate  
characterization of the physical state of the ICM gass and to assess the relative 
strength of cosmological effects shaping the evolution of galaxy clusters.

In this work, we therefore investigate, for the first time, the effect that primordial non-gaussianities have 
on galaxy clusters scaling relations, using hydrodynamic N-body simulations of large scale structure. 
We focus on scalings involving cluster mass, $M$, and gas 
properties related to the thermodynamical state of the intra-cluster medium. These are the temperature, $T$, entropy, $S$ and the cluster integrated pressure (thermal energy density) expressed by the SZ 
$Y$-Compton parameter.
Throughout the paper, and unless stated otherwise, we adopt a standard flat 
$\Lambda CDM$ cosmological model, with a Hubble constant, $H_{0}$, equal to $100h\,\mathrm{km/s}\,\mathnormal{Mpc}^{-1}$, with $h = 0.7$, fractional densities of matter and baryons today of $\Omega_{m} = 0.3$, $\Omega_{b} = 0.04$ respectively, a scalar spectral index, $n_{s}$ , equal to 0.96, and a power spectrum amplitude $A=2.1\times 10^{-9}$, so that $\sigma_{8} = 0.809$.

\section{Numerical Simulations and Catalogue Construction}
\label{sec:sinum}

To asses the impact that primordial non-Gaussianities on galaxy clusters scaling relations,  
we carried out hydrodynamic $N$-body simulations of large-scale structure with 
the publicly available Gadget-2 TreePM code \citep{2005MNRAS.364.1105S}, featuring adiabatic gas physics. 
The simulations initial conditions were generated with the 2LPT code \citep{2012PhRvD..85h3002S}, assuming periodic boundary conditions on a cubic volume with $L=250\, h^{-1} {\rm Mpc}$ on the side and populated with $N=2 \times 300^3 $ particles of baryon and dark matter.
The matter power spectrum transfer function was computed with the  CAMB code \citep{lewis:1999,lewis:2014} for the set of cosmological parameters adopted in the previous Section. The resulting baryon and dark matter particle masses in the simulations are $6.4\times10^{10}\,h^{-1} {\rm M_{\odot}}$ and  $4.2\times10^{9}\,h^{-1} {\rm M_{\odot}}$, respectively. 
The gravitational softening in physical coordinates was $30\,h^{-1} {\rm kpc}$. The initial conditions were generated  for different levels of non-gaussianity, allowing $f_{NL}$ to vary in the range $[-500,500]$ as indicated in Table~\ref{tab_runs}. For each value of $f_{NL}$, $5$ random box realizations were created with different seeds, thus resulting in a total of $35$ simulation runs. For each run, we have stored a total of 22 snapshots, with abutting boxes, in the redshift range $0\leq z \leq 2$.

\begin{table}
\begin{center}
\caption{\label{tab_runs} List of models considered in this work. Non-gaussian models are identified by the prefix ``NG'' followed by the corresponding $f_{NL}$ value, whereas ``G'' stands for the gaussian $\Lambda $-CDM model. Five initial condition realizations were produced for each model, yielding a total of 35 simulation runs. The quantities $N(z=0)$ and $N(z=1)$ are the total number of clusters when the five realizations for each model are combined. $\overline N(z=0)$ and $\overline N(z=1)$ give the average number of clusters for each realization of a given model.}
\begin{tabular}{ccrrrr}
\hline
   Model  & $f_{NL}$ & $N\left(z=0\right)$  & $N\left(z=1\right)$  & $\overline{N}\left(z=0\right)$    & $\overline{N}\left(z=1\right)$\\
\hline
NG-500  & -500 & 3460 & 771 & 692 & 142 \\
NG-300  & -300 & 3499 & 828 & 700 & 166 \\
NG-100  & -100 & 3649 & 892 & 730 & 178 \\
G       & 0    & 3653 & 913 & 731 & 183 \\
NG 100  & 100  & 3638 & 925 & 728 & 193 \\
NG 300  & 300  & 3713 & 1037 & 743 & 207 \\
NG 500  & 500  & 3779 & 1073 & 756 & 215 \\
\hline
\end{tabular}
\end{center}
\end{table}

To construct cluster catalogues for all runs, we used a  modified version of the cluster finder software developed by Thomas and collaborators  \citep{thomas:1998, pearce:2000, muanwong:2001}.  
The mass of the identified objects is set according to usual definition, 
\begin{equation}
M_\Delta(<R_\Delta)=\frac{4\pi}{3}R^3_\Delta\,\Delta\,\rho_{\rm crit}(z).
\label{eq:mass}
\end{equation}
where $\Delta$ is a fixed overdensity contrast, $\rho_{\rm crit}(z)=(3H_0^2/8\pi G)E^2(z)$ is the critical density and $E(z)=H(z)/H_0=\sqrt{(\Omega(1+z)^3+\Omega_\Lambda}$. 
Catalogue cluster properties are evaluated inside spheres of radius $R_{\Delta}$, centered around the densest dark matter particle in each cluster.
For this paper we chose $\Delta=200$ and set the minimum number of cluster particles equal to 500. In this way our original cluster catalogues are complete in mass down to $\approx3.41\times10^{13}h^{-1}M_{\sun}$, at all redshifts.  
For the present analysis, we trimmed our original catalogues to exclude galaxy groups with masses below $M_{\rm lim}= 5\times10^{13}h^{-1}M_{\sun}$. 
For each model, we also combined catalogues from different realization runs at each redshift to construct single cluster catalogues, all having a minimum mass limit, $M_{\rm lim}$.

Table~\ref{tab_runs} provides an overview of the number of clusters with masses above $M_{\rm lim}$ at $z=1$ and $z=0$ for each of our simulated models. The $N(z=0)$ and $N(z=1)$ are the total number of clusters when the five realizations of each model are combined. $\overline N(z=0)$ and $\overline N(z=1)$ give the average number of clusters for each realization of a given model.  
These numbers confirm expectations that cluster abundances are a function of $f_{NL}$, with negative/positive $f_{NL}$ models giving lower/higher cluster abundances than the gaussian model, see e.g. \cite{grossi:2007}.
Although our simulations were not set for mass function studies (they have a limited boxsize and five realizations for each model) we see that all our models follow this trend with the exception of the NG~100 model at $z=0$ that has the largest dispersion of initial conditions power spectrum amplitudes of all models.

Cluster properties investigated in this paper are the mass, $M$,
mass-weighted temperature, $T_{\rm mw}$, entropy, $S$ 
(defined as $S=k_{\rm B}T/n^{-2/3}$ where $T$ and $n$ are the gas temperature and number density), 
integrated Compton signal, $Y$ (defined as the
SZ signal times the square of the angular diameter distance to the
cluster), and $Y_{\rm X}$, the integrated Compton signal estimated using X-ray emission-weighted temperature, $T_{\rm X}$, and the gas mass, $M_{\rm gas}$. These quantities were computed using their usual definitions, see e.g. \citet{dasilva:2004}:	

\begin{equation}
M=\sum_k m_k,
\label{eqn:m}
\end{equation}
\begin{equation}
T_{\rm mw} = { \sum_{i} m_i \, T_i \over \sum_{i} m_i},
\label{eqn:tmw}
\end{equation}
\begin{equation}
S = { \sum_{i} m_i \, k_{\rm B}T_i\, n_i^{2/3} \over \sum_{i} m_i},
\label{eqn:s}
\end{equation}
\begin{equation}
Y = \frac{k_{\rm B}\sigma_{\rm T}}{m_{\rm e}c^2}\frac{(1+X)}{2m_{\rm H}}
\, \sum_{i}{m_i \, T_i}, 
\label{eq:y}
\end{equation}
\begin{equation}
Y_{\rm X} = \frac{k_{\rm B}\sigma_{\rm T}}{m_{\rm e}c^2}\frac{(1+X)}{2m_{\rm H}}
\, M_{\rm gas}\, T_{\rm X}
\end{equation}
\begin{equation}
M_{\rm gas} = \sum_i m_i,
\label{eqn:mgas}
\end{equation}
\begin{equation}
 T_{\rm X}=\frac{ \sum_{i} {m_i \, \rho_i \, \Lambda_{\rm bol}(T_i,Z)\, T_i}}
{\sum_{i} {m_i \, \rho_i \, \Lambda_{\rm bol}(T_i,Z)}} \,
\label{eqn:tx}
\end{equation}
where summations with the index {\it i} are over hot ($T_i > 10^5$K)
gas particles and the summation with the index {\it k} is over all
(baryon and dark matter) particles within $R_{200}$. Hot gas is
assumed fully ionized. The quantities $m_i$, $T_i$, $n_i$ and $\rho_i$
are the mass, temperature, number density and mass density of gas
particles, respectively. $\Lambda_{\rm bol}$ is the bolometric cooling
function in \citet{sutherland:1993} and $Z$ is the gas metallicity.
Other quantities are the Boltzmann constant, $k_{\rm B}$, the Thomson
cross-section, $\sigma_{{\rm T}}$, the electron mass at rest, $m_{{\rm
e}}$, the speed of light $c$, the Hydrogen mass fraction, $X=0.76$,
the gas mean molecular weight, $\mu $, and the Hydrogen atom mass,
$m_{\rm H}$.

\section{Scaling Relations}
\label{sec:scal}

In this paper we study the impact of non-gaussian models on galaxy cluster scaling relations of temperature, $T_{\rm mw}$,  entropy, $S$, and the $Y$ and $Y_{\rm X}$ SZ luminosities with the cluster mass, $M$. Following  \cite{dasilva:2009, aghanim:2009}, these scalings can be written as: 
\begin{equation}
T_{\rm mw}=A_{\rm TM}\,(M/M_0)^{\alpha_{\rm TM}}\,(1+z)^{\beta_{\rm TM}}\, E(z)^{2/3}  \,,
\label{eq:tm}
\end{equation}
\begin{equation}
S=A_{\rm SM}\,(M/M_0)^{\alpha_{\rm SM}}\,(1+z)^{\beta_{\rm SM}}\, E(z)^{-2/3} \,,
\label{eq:sm}
\end{equation}
\begin{equation}
Y=A_{\rm YT}\,(M/M_0)^{\alpha_{\rm YM}}\,(1+z)^{\beta_{\rm YM}}\, E(z)^{2/3} \,,
\label{eq:ym}
\end{equation}
\begin{equation}
Y_{\rm X}=A_{\rm YxM}\,(M/M_0)^{\alpha_{\rm YxM}}\,(1+z)^{\beta_{\rm YxM}}\, E(z)^{2/3} \,,
\label{eq:yxm}
\end{equation}
where $M_0$ was set equal to $10^{14}h^{-1}{\rm M_{\odot}}$ and all cluster properties are evaluated within $R_{200}$ (see Eq.~(\ref{eq:mass})). In this way, the redshift evolution of each scaling is modeled by a power of the $E(z)$ function, giving the predicted evolution extrapolated from the self-similar model \citep{kaiser:1986, kravtsov:2012}, times a power-law of $(1+z)$ accounting for departures to self-similar evolution. The
quantities, $A$, $\alpha$, and $\beta$, are therefore the scalings normalization at
$z=0$; the mass power-law index; and the index of the redshift power-law giving the deviation to self-similar evolution, respectively. Whenever $\beta=0$ the redshift evolution of the scalings is said be self-similar. Under the assumptions in  \cite{kaiser:1986}, the self-similar power-law indexes of the mass are $\alpha_{\rm TM}=\alpha_{\rm SM}=2/3$, and $\alpha_{\rm YM}=\alpha_{\rm YxM}=5/3$.

To determine $A$, $\alpha$, and $\beta$ for each scaling we use the
method described in \citet{dasilva:2004, aghanim:2009}. This involves re-writting Eqs~(\ref{eq:tm})--(\ref{eq:yxm}) in a logaritmic, concise, form,
\begin{eqnarray}
& &\log (y\, f(z))=\log (y_0(z)) + \alpha \, \log(x/x_0)  \,, 
\label{eqyx}\\
& &\log (y_0(z)) = \log (A) + \beta \, \log(1+z) \,,
\label{eqy0z}
\end{eqnarray}
where $y$ and $x$ are cluster properties, and $f(z)$ is some fixed power of the cosmological factor $E(z)$.
The method starts with a fit of the cluster populations at each redshift with Eq.~(\ref{eqyx}). 
If the logarithmic slope $\alpha$ does not change (i.e. shows no systematic variations) with $z$, the fitting procedure is then repeated with $\alpha$ set to its value at redshift zero, 
$\alpha(z=0)$, and the scaling normalisation factors $y_0(z)$ are stored. 
In this way we avoid unwanted correlations between $\alpha$ and the normalizations $y_0(z)$.
At this step we also store the r.m.s. dispersion of the fits at each redshift, 
\begin{equation}
\sigma_{\log y'} =\sqrt{ {1 \over N} \sum_{i} (\log (y'_i/y'))^2} \,,
\label{sigmay}
\end{equation}
where $y'=yf$ (see Eq.~(\ref{eqyx})) and $y'_i$ are individual data
points. 
To determine the parameters $A$ and $\beta$, we fit Eq.~(\ref{eqy0z}) to the stored values of $\log (y_0(z))$ as a function of $\log (1+z)$.
Since cluster abundances drop rapidly with $z$ (see Table~\ref{tab_runs}), we limited the present cluster scaling analysis to the redshift range $0 \le z \le 1 $, so that the fitting procedure is carried out with a reasonable number of clusters for all realization runs.  
We have also checked that the application of this fitting procedure to individual realization catalogues and to single catalogues that combine clusters from realizations runs of each model lead to equivalent results for the derived scalings. We therefore use the latter catalogues to display fitting values and figures, from this point onwards.

\section{Results}

\subsection{Scaling relations at redshift zero}

\begin{figure*}
\includegraphics[scale=0.6]{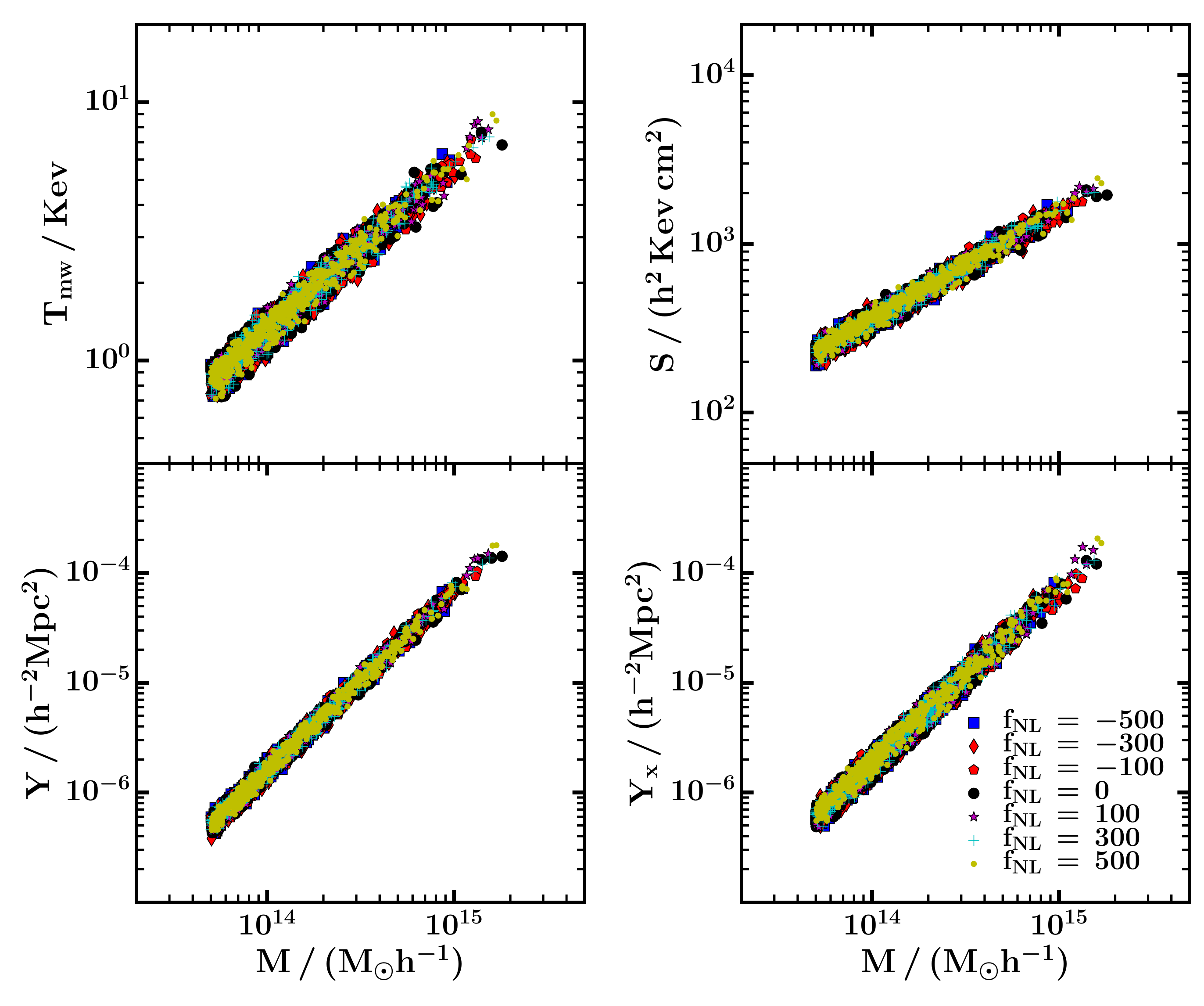}
\caption{Cluster scalings at redshift zero for the $T_{mw}-M$ (top left panel), $S-M$ (top right panel),  $Y-M$ (bottom left panel) and $Y_{\rm X}-M$ (bottom right panel), for values of $f_{NL}$ ranging from $-500$ to $500$ with increments of $200$ and a guassian model, $f_{NL}=0$ . The displayed quantities were computed within $R_{200}$. 
For clarity, we only plotted $500$ clusters randomly selected from the catalogues for each model. 
}
\label{fig:Scaling-Relations}
\end{figure*}

In this Section we discuss the scalings Eqs~(\ref{eq:tm})--(\ref{eq:yxm}) obtained at redshift zero, from our suite of N-body/hydrodynamic simulations runs with non-gaussian initial conditions.

Figure~\ref{fig:Scaling-Relations} shows the galaxy cluster distributions for the scalings: $T_{mw}-M$ (top left), $S-M$ (top right), $Y-M$ (bottom left) and $Y_{\rm X}-M$ (bottom right), with quantities computed within $R_{200}$. In each panel,  models are labeled according to their values of $f_{NL}$: $-500$ blue squares, $-300$ red diamonds, $-100$ red pentagons, $0$ black filled circles, $100$ magenta asterisks, $300$ cyan pluses and $500$ yellow dots. To improve clarity, we only display 500 clusters for each model, randomly drawn from the combined realizations catalogues with a weighting procedure that guarantees that the most massive and rare objects are displayed.

A common trend in all panels is that $T_{mw}$, $S$, $Y$ and $Y_{\rm X}$ are properties tightly related to the total cluster mass. This confirms expectations, because temperature is weighted by mass (not by X-ray emission) and entropy is computed using $T_{\rm mw}$ which is a better proxy than $T_{\rm X}$ for the thermodynamic temperature. On the other hand, the cluster integrated SZ signal is a measure of the total thermal energy of the object, which is known to be more dependent on the cluster total gravitational mass and gas mass fraction than on the details of gas physical effects acting inside $R_{200}$. The $Y_{\rm X}-M$ relation displays larger dispersions than the $Y-M$ scaling, because the former is 
computed using the X-ray emission-weighted  temperature, $T_{\rm X}$, which is more sensitive to internal gas physical effects than $T_{\rm mw}$. 

Table~\ref{tab_best_fit} presents the best fit parameters,  $\alpha (z=0)$ and $\log A$, and fit dispersions, 
$\sigma_{\log y'} \left[z=0\right]$, of our cluster scaling relations at redshift zero (see Section~\ref{sec:scal}).
In general, all scalings show very similar slopes for the various models. Low $f_{NL}$ models seem to have slightly smaller slopes but variations are consistent within one to two $1-\sigma$ errors giving the statistical uncertainties of the fits.
The results for $\log A$ in Table~\ref{tab_best_fit}, indicate that the normalization of the scalings at $z=0$ has a mild but systematic increase with $f_{NL}$. This is impossible to visualize in the plots of each scaling due to the intrinsic dispersions of the fits. 
Finally all $z=0$ scalings show fit dispersions $\sigma_{\log y'} \left[z=0\right]$ which are independent of the level of primordial non-gaussianities. 
According to Table~\ref{tab_best_fit} the intrinsic dispersion of the $Y_{\rm X}-M$ scaling is about 1.8 times larger than the dispersion of the $Y-M$ scalingat $z=0$.

\subsection{Evolution of Scaling Relations \label{scale_evol}}

\begin{figure*}
\includegraphics[scale=0.35]{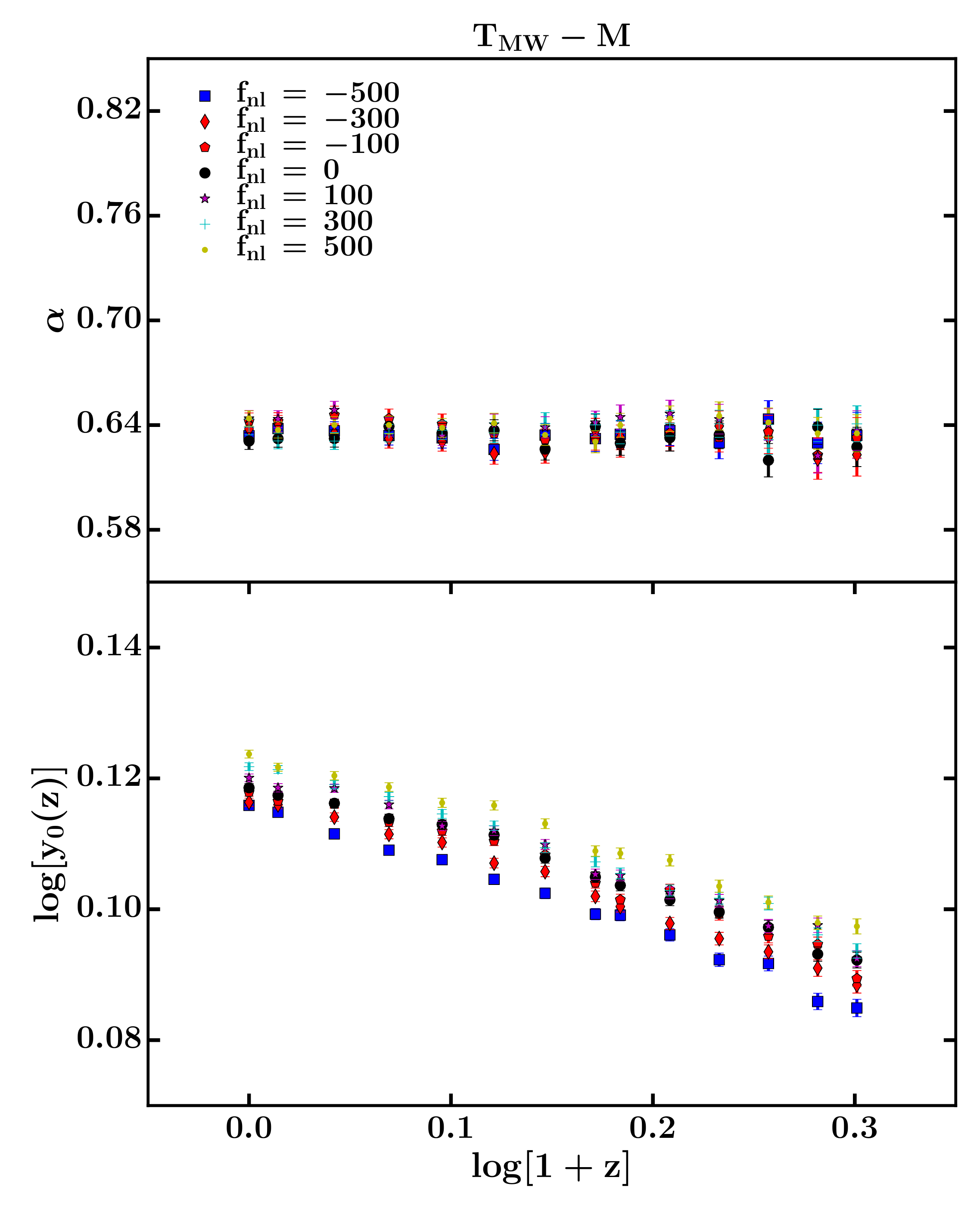}
\includegraphics[scale=0.35]{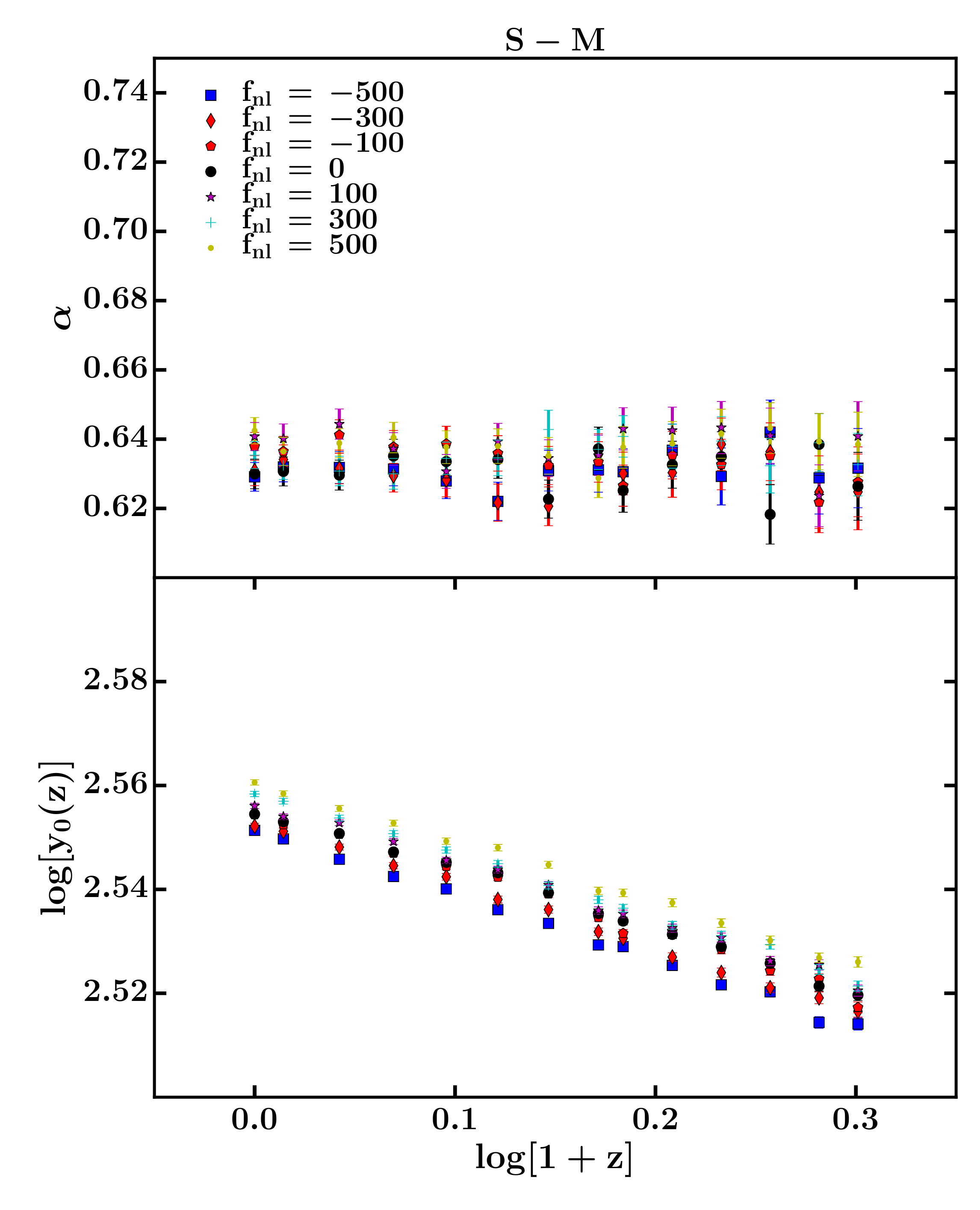}\\
\includegraphics[scale=0.35]{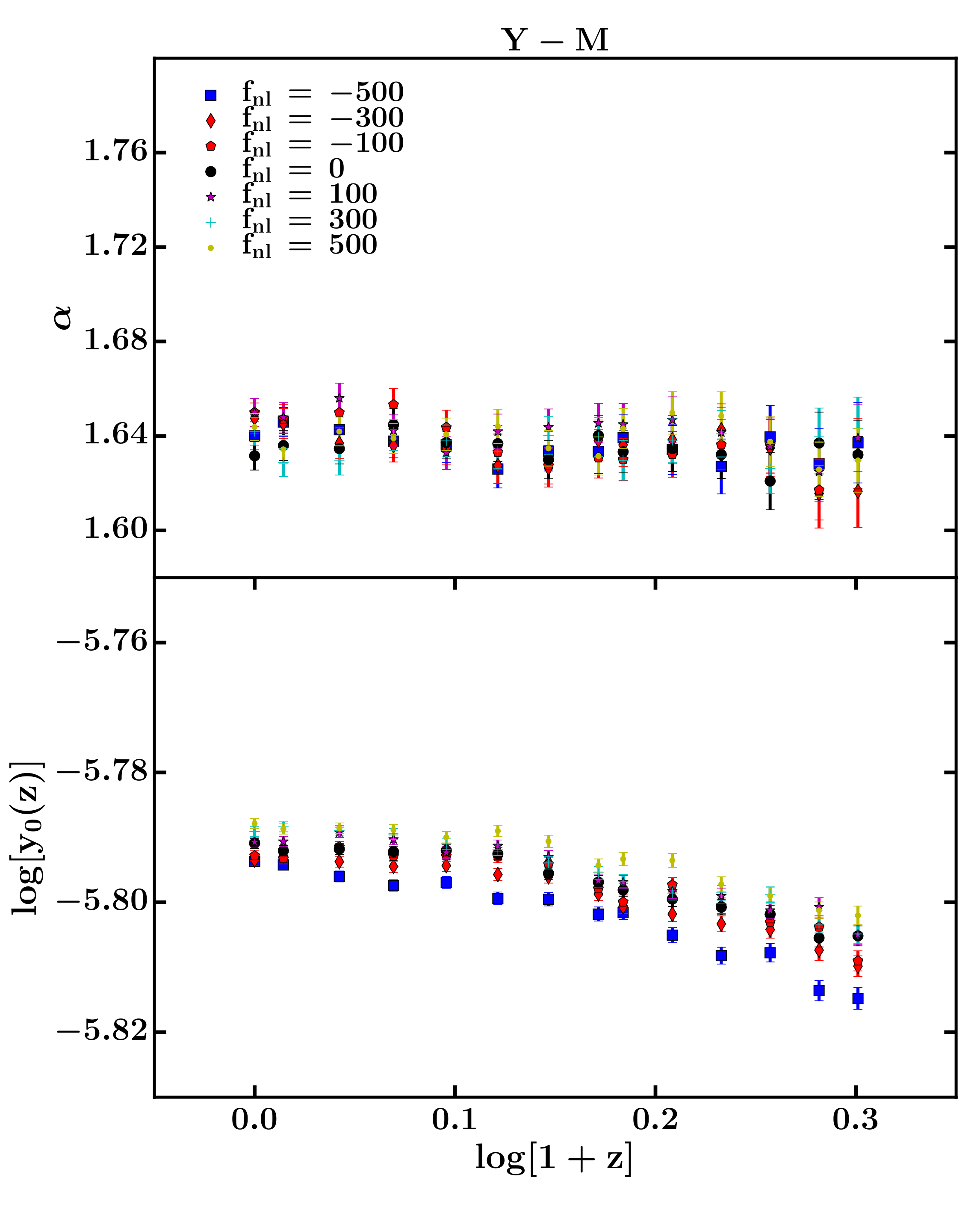}
\includegraphics[scale=0.35]{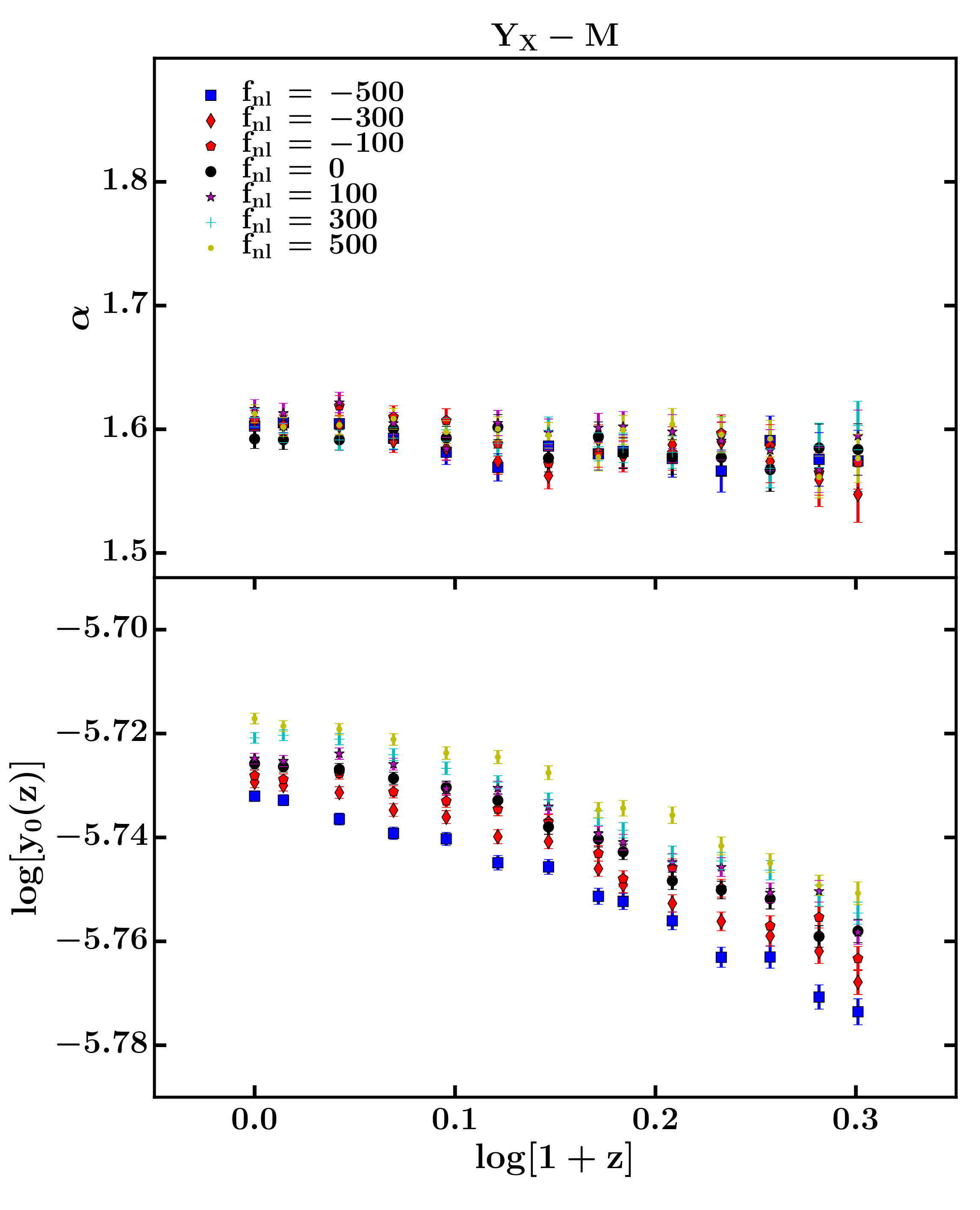}

\caption{The evolution of the slope, $\alpha$, normalization, $\log_{10}\left(\left(z\right)\right]$ and respective $1-\sigma$ error bars,  with redshift for $T_{mw}-M$ (top left panel), $Y-M$, (top right panel), $S-M$ (bottom left panel) and $L_{X}-M$ (bottom right panel), for different values of $f_{NL}$ ranging from $-500$ to $500$ with increments of $100$.   }
\label{fig:scale_evol}
\end{figure*}

To study the evolution of the scaling laws we applied the method described in Section~\ref{sec:scal} to the full set of cluster catalogues in our simulations. As mentioned earlier, we carried out the analysis in two ways. One applies the method to catalogues from individual realization runs, from which averaged fitting parameters were inferred for each model. A second approach consisted in combining individual realization catalogues at each redshift and then applying the fitting procedure to the resulting combined catalogues to obtain the scaling parameters. We verified that both approaches lead to equivalent scaling parameters within the defined range of redshifts, $0\le z\le 1$. The results presented in this paper are from the second approach, which somewhat simplifies the presentation of results and the legibility of plots.

The main result of this Section is the set of plots presented in Figure~\ref{fig:scale_evol}. These show the evolution of the fitting parameters in 
(Eq.~\ref{eqyx}), the power-law index $\alpha $ and the normalization 
$\log (y_0(z))$, for all scalings and models considered in this paper.
The figure is divided in four plots, one for each scaling (top left: $T_{mw}-M$; top right: $Y-M$; bottom left: $S-M$; bottom right: $Y_{\rm X}-M$). Each plot contains two panels displaying the evolution of $\alpha $ (top panel) and $\log (y_0(z))$ (bottom panel) with $z$. Models are labeled in the same way as in 
Fig.~\ref{fig:Scaling-Relations} and bars in data the points are $1-\sigma$ bootstrap resampling errors.

A first conclusion from Fig.~\ref{fig:scale_evol} is that the power-law index, 
$\alpha $ shows no systematic variation with $z$ for all scalings. In general, data points and errors appear scattered around the redshift zero value, 
$\alpha (z=0)$, for each $f_{\rm NL}$ model in all scalings. 
We note that although our simulations include only adiabatic gas physics all $\alpha$ points (including those from the Gaussean $\Lambda$ model) are, in general, below the self-similar predictions: $\alpha_{\rm TM}=\alpha_{\rm SM}=2/3$, and $\alpha_{\rm YM}=\alpha_{\rm YxM}=5/3$. These predictions assume hypothesis such as hydrostatic equilibrium and spherical symmetry in clusters (as well as a critical density cosmology, \citep{kaiser:1986}) which are only approximations to the true state of clusters in simulations \citep{kravtsov:2012}.
Deviations from self-similar values are small but in most cases larger than the statistical errors. The larger deviations are found for the $Y_{\rm X}-M$ scaling, which presents systematically lower $\alpha$ than the $Y-M$ scaling. This is because the SZ signal is proportional to the product of the cluster gas temperature by mass ($Y_{\rm SZ}\propto TM$) and the temperature scales in our simulations as 
$T_{\rm mw} \propto M^{0.64}$ and $T_{\rm X} \propto M^{0.60}$ (these values are good approximations for all models). In this paper we will not display further results for the $T_{\rm X}-M$ scaling, which has an evolution for the gaussian model 
consistent with the results in Fig.~2 for the $w=-1$ simulations in \cite{aghanim:2009} (their simulations have a smaller boxsize but the same gas physics and similar cosmology to our G model runs).

The scaling-law normalizations, $\log (y_0(z))$, in Fig.~\ref{fig:scale_evol} denote clear trends with redshift and $f_{\rm NL}$.  
The decrease of $\log (y_0(z))$ with $z$ puts in evidence that all scalings tend to deviate from self-similar evolution, in a way that clusters of a given mass have lower temperatures, entropies and $Y_{\rm SZ}$ signals at higher $z$ than what would be expected assuming self-similar evolution. The panels show that this negative (with respect to self-similar) evolution follows, in general, linear trends with $z$ that can be fit with Eq.~(\ref{eqy0z}) using the method described in Section~\ref{sec:scal}.
Table~\ref{tab_best_fit}, lists the normalization constant, $A$ and the power-law index $\beta$ modeling the redshift dependence of $\log (y_0(z))$ obtained in this way for all scalings. These numbers confirm negative $\beta$ slopes with mild (but statistically significant) deviations from the self-similar expectation $\beta=0$. 
The dependence of the $\log (y_0(z))$ normalization with  $f_{\rm NL}$ is also %quite 
evident from Fig.~\ref{fig:scale_evol}. 
For each scaling, models with higher  $f_{\rm NL}$ tend to show larger normalizations at all redshifts. This can be understood in light of the findings in N-body simulations \citep{smith:2011} and analytical modeling using excursion set theory \citep{dizgah:2013} that cluster haloes in non-gaussian models have increased/decreased core densities for positive/negative $f_{NL}$. As a consequence cluster gas properties such as temperature, entropy and the$Y_{\rm SZ}$ signal are expected to follow this trend, leading to scaling normalizations that increase with $f_{\rm NL}$. 

An interesting aspect to address with cluster simulations is to investigate the evolution of the intrinsic scatter of scaling laws with redshift. In our simulations we find that cluster scaling laws involving mass-weighted quantities (i.e. the $T_{mw}-M$, $S-M$ and $Y-M$ scalings) show no significant evolution of fit dispersions, $\sigma_{\log y'}$, with redshift. The quantities  $\sigma_{\log y'} \left[z=0\right]$ and $\sigma_{\log y'} \left[z=1\right]$ in Table~\ref{tab_best_fit} give the fit dispersions in our models at $z=0$ and $z=1$, respectively. 
For the $Y_{\rm X}-M$ scaling our simulations indicate an increase of the fit dispersions with z. This effect is independent of $f_{\rm NL}$ and is related to the fact that $Y_{\rm X}$ depends on $T_{\rm X}$, which in turn is a function the evolution of gas X-ray emission with redshift.

\subsection{Dependence on $f_{\rm NL}$}

\begin{figure*}
\includegraphics[scale=0.35]{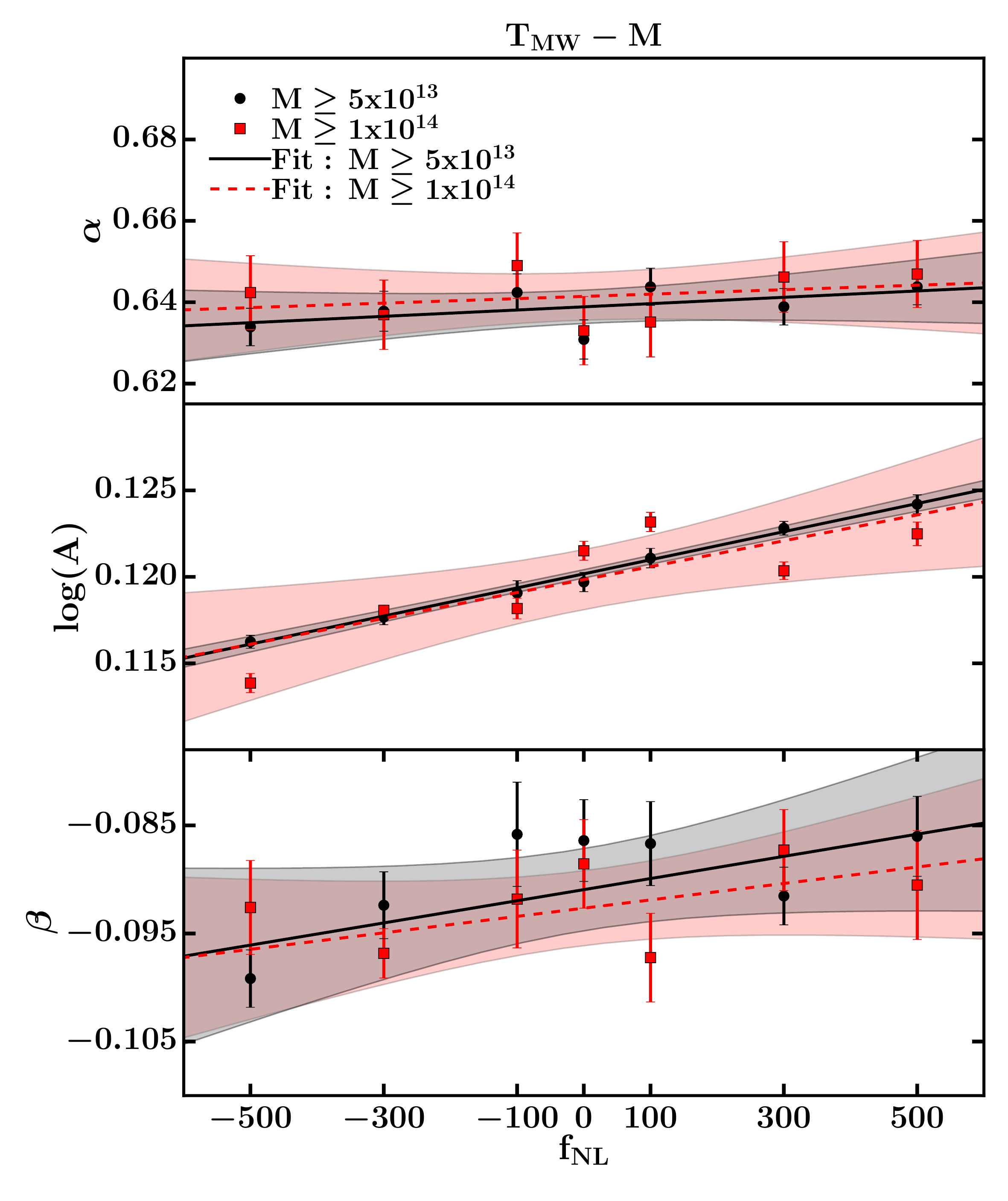}
\includegraphics[scale=0.35]{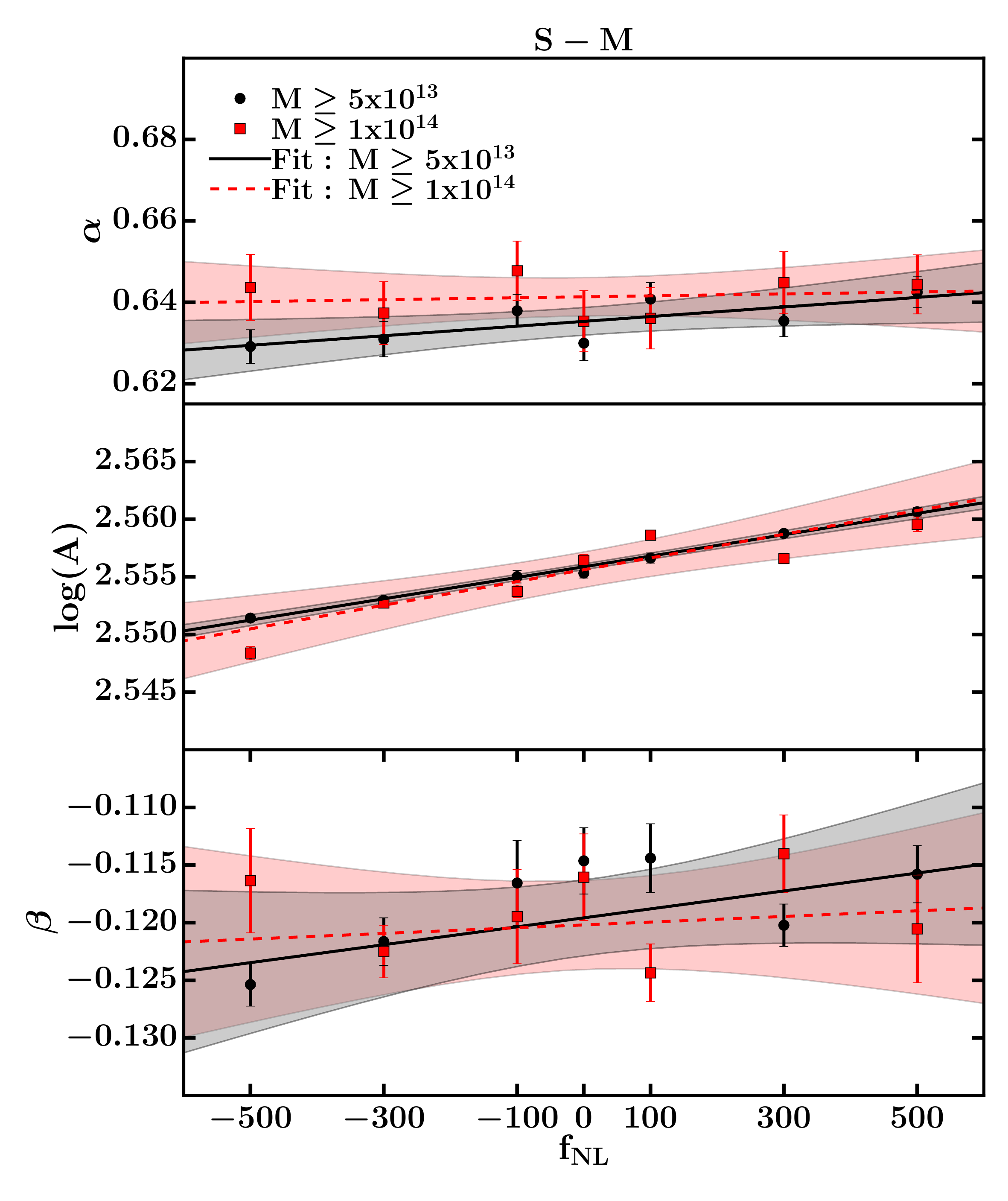}\\

\includegraphics[scale=0.35]{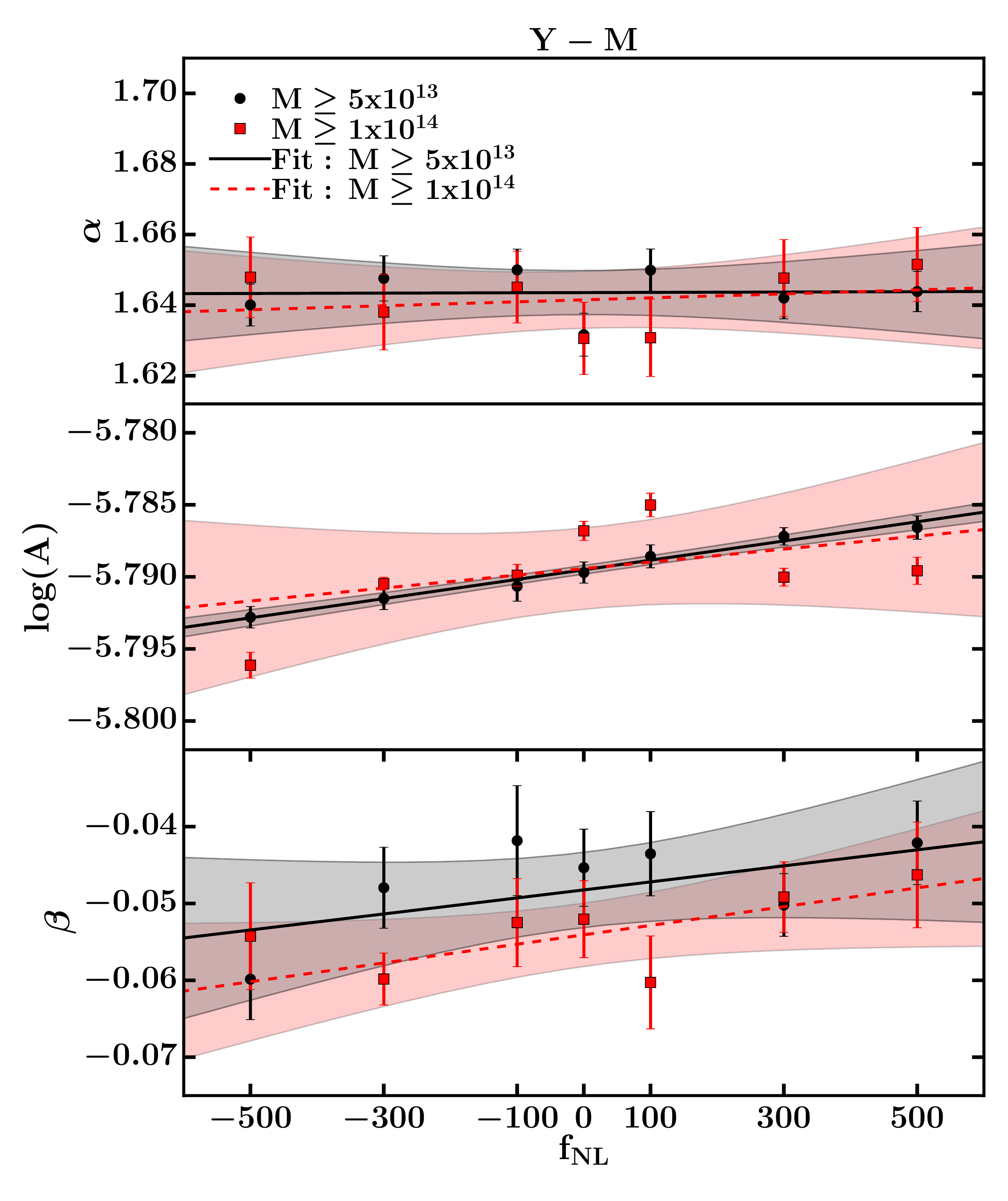}
\includegraphics[scale=0.35]{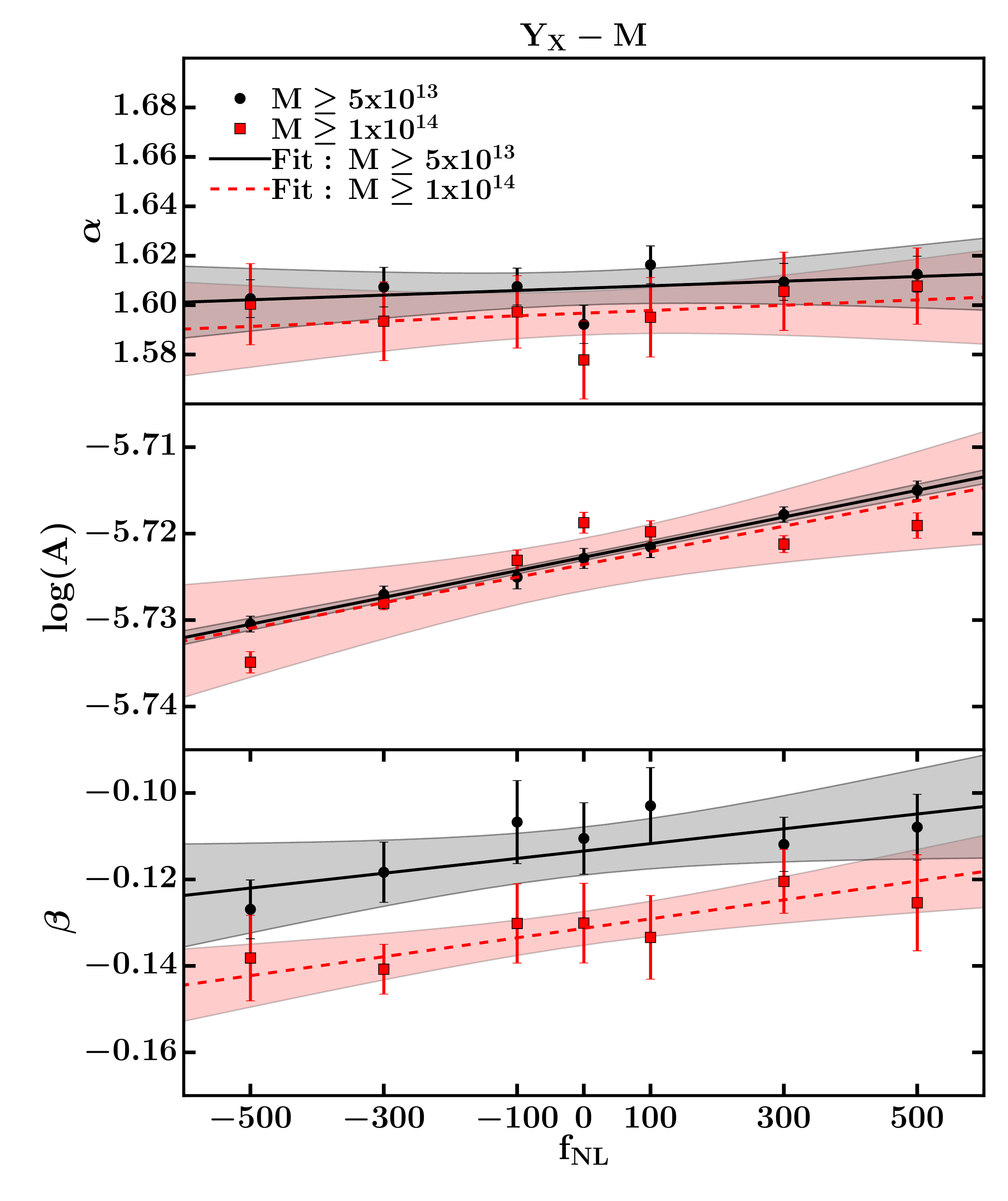}

\caption{The dependence of the power-law index $\alpha$ of mass, normalization parameter $\log_{10}\left(A \right)$, and power-law index of redshift $\beta$ and their respective $1-\sigma$ error bars, as a function of $f_{NL}$  for $T_{mw}-M$ (top left panel), $Y-M$, (top right panel), $S-M$ (bottom left panel) and $Y_{\rm X}-M$ (bottom right panel). Black solid line and black shaded area corresponds to the linear fit ans $95\%$ C.L. confidence interval for a mass cut of $5\times10^{13}\,M_{\odot}h^{-1}$, while dashed red line and red shaded are corresponds to the linear fit ans $95\%$ C.L. confidence interval for a mass cut of $1\times10^{14}\,M_{\odot}h^{-1}$. 
\label{fig:scale_fnl}
}
\end{figure*}

With the results from Table~\ref{tab_best_fit}  we constructed plots in Fig.~\ref{fig:scale_fnl} that put in evidence the impact of non-gaussian initial conditions on the four galaxy cluster scaling-laws investigated in this paper.
The panels in each plot give the best fit values for the mass power-law index $\alpha $ (top panel), the scaling normalization $\log A$ (centre panel), and the $(1+z)$ power-law index $\beta $ (bottom panel) as a function of $f_{\rm NL}$. Data in black are the results from Table~\ref{tab_best_fit}.
To test the robustness of the results with respect to a different choice of $M_{\rm lim}$ we repeated the analysis in the previous sections imposing a higher minimum mass limit, $M_{\rm lim}=1\times 10^{14} h^{-1}M_{\sun}$, to our simulated catalogues.
This analysis leads to the data displayed in red.
For both colour-coded data sets, bars indicate bootstrap errors, lines are straight-line fits to the data points, and shaded areas represent the $95\%$ confidence levels preferred by the data.

These plots indicate that the mass power-law index $\alpha $ remains approximately unchanged with $f_{\rm NL}$. 
Variations are as small as 1.9\%, 1.3\%, 0.6\% and 0.1\% for the scaling $S-M$, $T_{mw}-M$, $Y_{\rm X}-M$ and $Y-M$, respectively.
When less massive clusters and groups are excluded from the analysis (see data points from the $M_{\rm lim}=1\times 10^{14} h^{-1}M_{\sun}$ catalogues), the dependence of $\alpha $ on $f_{\rm NL}$  is even weaker for $T_{mw}-M$, $S-M$ and $Y-M$ scalings, with variations of about 0.9\%, 0.3\% and 0.3\% respectively; while for $Y_{\rm X}-M$ scalings the dependence is slightly stronger but not larger than 1\%. This means that the $\alpha $ variations with $f_{\rm NL}$ in our $M_{\rm lim}=1\times 10^{14} h^{-1}M_{\sun}$ catalogues are always below the one percent level for all scalings.

The scaling laws normalization parameter $A$ is slightly more sensitive to non-Gaussianities. 
Within the displayed range of $f_{\rm NL}$, the normalization parameters $A$ change by about 3.8\% for $Y_{\rm X}-M$, 2.1\% for $S-M$, 1.9\% for $T_{\rm mw}-M$ and 1.6\% for $Y-M$ scalings. Similar variations are found for the results obtained with $M_{\rm lim}=1\times 10^{14} h^{-1}M_{\sun}$ catalogues.

The impact of non-gaussian initial conditions is stronger for the redshift power-law index, $\beta$, that measures the departures from self-similar evolution of the scalings.  
The variations of $\beta$ within the displayed range of  $f_{\rm NL}$ are about 20\%, 13\%, 11.4\% and 6.5\% for the scalings $Y-M$, $Y_{\rm X}-M$, $T_{mw}-M$ and $S-M$, respectively.
When the less massive objects are excluded from the analysis (catalogues with $M_{\rm lim}=1\times 10^{14} h^{-1}M_{\sun}$), the $T_{mw}-M$ and $S-M$ scalings show weaker variations with $f_{\rm NL}$. The SZ scalings show slightly larger percentage variations but systematically lower $\beta $ when compared with the results from the $M_{\rm lim}=5\times 10^{13} h^{-1}M_{\sun}$ catalogues.  

The effect of non-gaussian initial conditions on these cluster scalings is consistent with the view that positive/negative $f_{\rm NL}$ tend to increase/decrease cluster concentrations \citep{smith:2011}. 
Clusters with higher concentrations
tend to have higher gas densities and temperatures (and therefore higher entropy and the $Y_{\rm SZ}$ signal) at the their inner regions. According to our findings, this influences the normalization $A$ and the evolution $\beta$ parameters of the cluster scaling laws.
We note that, although $f_{\rm NL}$ has a significant impact on $\beta $, these departures from self-similar evolution are in general small for all scalings.
According to \citep{smith:2011} the effect of non-gausseanity on cluster concentrations increases slightly with mass. 
This effect appears not to have a too strong impact on the cluster fitting parameters when we change the minimum mass limit of our catalogues to $M_{\rm lim}=1\times 10^{14} h^{-1}M_{\sun}$. 
The exception may be the $\beta $ parameters in the $Y_{\rm SZ}$-mass scalings, which show a slight increase when low-mass clusters and groups are excluded from the catalogues. This tendency is however reversed in the case of the $T_{mw}-M$ and $S-M$ scalings. 
We note, however, that the effect of cluster concentrations is in competition with other effects such as the increase of scatter due to a reduction of the total number of clusters in the fitting procedure when the minimum mass limit of the catalogues is increased to $M_{\rm lim}=1\times 10^{14} h^{-1}M_{\sun}$.

\begin{table*}
\begin{center}
\caption{\label{tab_best_fit} Best fit values of the parameters
  $\alpha$, $\log A$ and $\beta$ as well as their respective $1\sigma$
  errors. These values are valid within the redshift range $0\le z\le1$.}
\begin{tabular}{l|rrrrrrr}
\hline
     & Model NG-500 & Model NG-300 & Model NG-100 & Model G & Model NG100 & Model NG300 & Model NG500  \\
\\
\hline

$T_{MW} - M $ & &  &   &   &  &  & \\
$\alpha\left(z=0\right)$  & $0.634\pm0.005$ & $0.638\pm0.005$ & $0.642\pm0.005$   & $0.631\pm0.005$  &$0.644\pm0.005$  & $0.639\pm0.005$  & $0.644\pm0.004$\\
 $log_{10}\left(A\right)$ & $0.116\pm0.001$ & $0.118\pm 0.001$ & $0.119\pm 0.001$   & $0.120\pm 0.001$  &$0.121	\pm 0.001$  & $0.123\pm 0.001$  & $0.124\pm	 0.001$\\
$\beta$  & $-0.099\pm 0.003$ & $-0.092\pm 0.003$ & $-0.086\pm 0.005$   & $-0.086\pm 0.004$  &$-0.087\pm 0.004$  & $-0.092\pm 0.003$  & $-0.086\pm	 0.004$\\
$\sigma_{\log y^{\prime}}\left[z=0\right]$ & $0.0012$ & $0.0014$ &$0.0013$   &  $0.0014$ & $0.0013$  & $0.0013$ & $0.0013$\\
$\sigma_{\log y^{\prime}}\left[z=1\right]$ & $0.0012$ & $0.0013$  & $0.0013$  & $0.0013$  & $0.0014$ & $0.0013$  & $0.0014$ \\
\\
\hline
$S - M  $ & &  &   &   &  &  & \\
$\alpha\left(z=0\right)$  & $0.629\pm0.004$ & $0.631\pm0.004$ & $0.638\pm0.004$   & $0.630\pm0.004$ &$0.641\pm0.004$  & $0.635\pm0.004$ & $0.643\pm0.004$\\
 $log_{10}\left(A\right)$ & $2.551\pm 0.001$ & $2.553\pm 0.001$ & $2.555\pm 0.001$  & $2.555\pm 0.001$ &$2.557\pm 0.001$  & $2.559\pm 0.001$ & $2.561\pm	 0.001$\\
$\beta$  & $-0.125\pm 0.002$  & $-0.122\pm 0.002$ & $-0.117\pm 0.004$  & $-0.115\pm 0.003$ &$-0.114	\pm 0.003$  & $-0.120\pm 0.002$  & $-0.116\pm0.003$\\
$\sigma_{\log y^{\prime}}\left[z=0\right]$ & $0.0010$ & $0.0011$  &$0.0010$  &  $0.0011$ & $0.0010$  & $0.0010$  & $0.0010$\\
$\sigma_{\log y^{\prime}}\left[z=1\right]$ & $0.0009$ & $0.0010$   & $0.0010$ & $0.0010$  & $0.0011$ & $0.0010$  & $0.0011$ \\
\\
\hline
$Y - M $ & &  &   &   &  &  & \\
$\alpha\left(z=0\right)$  & $1.640\pm0.006$ & $1.648\pm0.006$ & $1.650\pm0.006$   & $1.632\pm0.006$  &$1.650\pm0.006$  & $1.642\pm0.006$ & $1.644\pm0.006$\\
 $log_{10}\left(A\right)$ & $-5.793\pm 0.001$ & $-5.792\pm 0.001$ & $-5.791\pm 0.001$  & $-5.790\pm 0.001$  &$-5.789\pm 0.001$  & $-5.787\pm 0.001$ & $-5.787\pm 0.001$\\
$\beta$  & $-0.060\pm 0.005$ & $-0.048\pm 0.005$ & $-0.042\pm 0.007$   & $-0.045\pm 0.005$  &$-0.044\pm 0.006$  & $-0.050\pm 0.004$H & $-0.042\pm0.005$\\
$\sigma_{\log y^{\prime}}\left[z=0\right]$ & $0.0020$& $0.0022$ & $0.0021$  &  $0.0021$ & $0.0021$ & $0.0021$ & $0.0022$\\
$\sigma_{\log y^{\prime}}\left[z=1\right]$ &$0.0020$ & $0.0021$ & $0.0021$  & $0.0021$  & $0.0022$ & $0.0020$ & $0.0020$\\
\\
\hline
$Y_{x} - M $ & &  &   &   &  &  & \\
$\alpha\left(z=0\right)$  & $1.603\pm0.008$ & $1.607\pm0.008$ & $1.608\pm0.008$  & $1.592\pm0.008$  &$1.616\pm0.008$  & $1.609\pm0.008$ & $1.613\pm0.007$\\
 $log_{10}\left(A\right)$ & $-5.731\pm 0.001$ & $-5.727\pm 0.001$ & $-5.725\pm 0.001$ & $-5.723\pm 0.001$ &$-5.722\pm 0.001$  & $-5.718\pm 0.001$ & $-5.715\pm 0.001$\\
$\beta$  & $-0.127\pm 0.007$ & $-0.118\pm 0.007$ & $-0.107\pm 0.01$  & $-0.111\pm 0.008$  &$-0.103\pm 0.009$  & $-0.112\pm 0.006$ & $-0.108\pm0.008$\\
$\sigma_{\log y^{\prime}}\left[z=0\right]$ & $0.0035$& $0.0038$ & $0.0038$  & $0.0039$  & $0.0038$ & $0.0039$ & $0.0039$\\
$\sigma_{\log y^{\prime}}\left[z=1\right]$ &$0.0044$ & $0.0046$ & $0.0048$  &  $0.0045$ & $0.0051$ & $0.0046$ & $0.0050$\\
\\
\hline
\end{tabular}
\end{center}
\end{table*}

\section{Summary}

In this paper we present galaxy cluster scaling relations from hydrodynamic/N-body simulations of large-scale structure, featuring adiabatic gas physics and non-gaussian initial conditions for the mater density fluctuations. 
We investigated five non-gaussian models with local $f_{\rm NL}$ parameterizations ranging from -500 to 500 
and a $f_{\rm NL}=0$ gaussian model, with a flat $\Lambda CDM$  cosmology. We did a total of 35 simulation runs and generated catalogues with cluster masses larger than $M_{\rm lim}=5\times 10^{13}\, h^{-1}M_{\sun}$ to study scaling relations involving mass-weighted temperature, $T_{\rm mw}$,  entropy, $S$, integrated SZ signals, $Y$ and $Y_{\rm X}$, with mass, $M$  (see Eqs~(\ref{eq:tm})--(\ref{eq:yxm})).

The main conclusions of this study are:
\begin{itemize}

\item Non-gaussian initial conditions have a mild but significant impact on the normalization of the cluster scalings, $y_0(z)$, and almost no impact on the power-law index, $\alpha $, of the mass dependence.
\item The normalizations $y_0(z)$ are affected by non-gaussianities through changes in the amplitude parameter $A$, giving the normalization of the scalings at $z=0$, and through variations in their non self-similar evolutions, parametrized by power-laws of $(1+z)$ with indexes $\beta$.
\item The redshift zero normalizations, $A$, show only slow increases with $f_{\rm NL}$, of the order of $1.6\%-3.8$\% in the range $-500 \le f_{\rm NL} \le 500$, for the various scalings. 

\item Non-gassianities have a stronger impact on the redshift evolution of the normalizations. Our $\beta$ parameters increase with $f_{\rm NL}$ by a maximum of $20\%$ for the $Y-M$ scaling and a minimum of $7\%$ for the $S-M$ scaling within $-500 \le f_{\rm NL} \le 500$. In all cases the $\beta $ parameters,
that measure departures from self-similar evolution, are found to be close to the  expected self-similar evolution of each scaling. 

\item Increasing the minimum mass limit of our catalogues to $M_{\rm lim}=1\times 10^{14}\, h^{-1}M_{\sun}$, we find similar dependences for $A$ and $\alpha$ with  $f_{\rm NL}$. The dependence of $\beta $ with $f_{\rm NL}$ becomes stronger for $Y_{\rm X}-M$ and $Y-M$ and less prominent  for the $T_{\rm mw}-M$ and $S-M$ scalings. 

\end{itemize}

These results are in line with the predictions that $f_{\rm NL}$ changes the internal structure 
of cluster profiles, as a result of an increase/decrease of cluster concentrations for 
positive/negative $f_{\rm NL}$ \citep{smith:2011, dizgah:2013}. The impact on cluster 
scaling relations is mostly due to changes in the evolution of their normalizations. Our results show 
that this impact is small for models with low $f_{\rm NL}$. 
However, for larger values of $f_{\rm NL}$, the effect of PNG on the evolution of cluster 
scalings can be as important as the effect of non-gravitational gas physics 
(see eg \cite{kay:2007, dasilva:2009}) or the effect of  
dark energy (see, eg \cite{aghanim:2009}) in clusters scaling relations.
This means that it is safe to neglect the effects of PNG on the investigated clusters scalings 
if the present observational constraints from Planck on a scale invariant $f_{\rm NL}$ are 
valid at galaxy cluster scales. 
This may no longer be true if $f_{NL}$ is in fact a scale dependent parameter. 
In this case, $|f_{NL}|$ may have a larger amplitude at clusters scales and therefore our results 
show that galaxy cluster scalings are sensitive to primordial non-gaussianities and should 
be taken into consideration when assessing the constraining power of cluster surveys or when
using future galaxy cluster data to infer cosmological parameters.

\section{Acknowledgements}

AMMT was supported by the FCT/IDPASC grant contract SFRH/BD/51647/2011. AMMT, Antonio da Silva acknowledge financial support from project PTDC/FIS/111725/2009, funded by Fundac\~ao para a Ci\^encia e a Tecnologia.

\end{document}